\documentstyle[preprint,aps]{revtex}
\begin{document}
\draft
\tightenlines
\title{Superconductive Static Quantum Logic}
\author{Xin Xue}
\address{
Department of Natural Resource Sciences\\
Macdonald Campus of McGill University\\
Ste-Anne-de-Bellevue, Quebec, Canada H9X 3V9\\
{\rm E-mail: xkhz@musicb.mcgill.ca}
}
\author{Haiqing Wei}
\address{
Department of Physics, McGill University\\
Montreal, Quebec, Canada H3A 2T8\\
{\rm E-mail: dhw@physics.mcgill.ca}
}
\maketitle

\begin{abstract}
Superconducting rings with exactly $\Phi _0/2$ magnetic flux threading are
analogous of Ising spins having two degenerate states which can be used to
store binary information. When brought close these rings interact by means
of magnetic coupling. If the interactions are properly tailored, a system of
such superconducting rings can accomplish static quantum logic in the sense
that the states of the rings interpreted as Boolean variables satisfy
the desired logic relations when and only when the whole system is in the
ground state. Such static logic is essential to carry out the static
quantum computation [1,2].
\end{abstract}

\newpage

{\bf
A recent discovery of static quantum computation shows that a quantum system
with properly tailored many-body interactions can compute in the static
manner [1,2]. An interesting experimental issue is to practically construct
a static quantum computer. Although nanotechnology holds the promise to
realize the desired strongly interacting quantum system [1,2,3], it awaits
further advancement. At present, a pedagogy demonstration of static quantum
logic and computation using the state-of-the-art technology should be very
striking. In this letter it is proposed that superconducting rings
electrically isolated but magnetically coupled to each other are well suitable
for static quantum logic. The well-established technology [4,5] promises an
almost immediately practical implementation of the static quantum computer in
superconductive devices.
}

The behavior of an isolated superconducting ring under a magnetic field is
well understood. Generally a persistent current develops in the ring to
satisfy the flux quantization condition [4]. When the flux due to the external
field threading the ring is $\Phi _0/2=(hc/2e)/2$, two directions of
supercurrent are energetically degenerate which are analogous to the two
possible orientations of an Ising spin. Later in this letter such a ring will
be called the Ising-spin-ring (ISR). If two ISRs are placed close, there is a
further analogy of ``antiferromagnetic'' (AFM) coupling [6]. The two states of
the ISR can be used to store binary information and static quantum logic
can be realized by tailoring the AFM coupling among the rings. Starting with
the ISRs, one may implement the binary wire and then other static quantum
logic gates [2] using the same idea as in references [1,2,3,7]. However the
convenience of conveying the supercurrent as information carrier through
extended superconducting wires makes it possible to realize static logic in
another way. The ISR under consideration in this letter is actually the 
double-ring consisting of two rings (called head and end respectively)
connected by two narrowly separated lines as shown in Fig.1. All ISRs have
the same total geometry area so that when subject to a suitable magnetic
field the flux threading all the double-rings should be exactly $\Phi _0/2$.
The two wises of circulating current are degenerate and representing logic $0$
and $1$ respectively. The convention using here is clock-wise supercurrent
for logic $0$ and the opposite for $1$ (actually the logic values of $0$ and
$1$ are corresponding to the numbers $0$ and $1$ of trapped flux quanta in the
ring). The advantage of using double-rings to implement the superconductive
static logic is that the logic gates can be made far enough from each other to
avoid the inter-gate interactions. By this way a double-ring with long
connecting wires can realize ``telecommunication'' conveniently.
Later in this letter such a double-ring will be simply called a cell.

As shown in Fig.2, with one cell for input and another one or more for
output, a logic inverter can be simply implemented by drawing the heads
of the output cells near the end of the input cell so that they are
antiferromagneticly interacting. Fig.2 also shows how to realize logic
fan-out. To construct a general purpose logic network, one needs nonlinear
logic gates like AND and OR gates. Fig.3 depicts a NAND/NOR gate which
carries both the NAND and the NOR operations [8]. There are small bias fields
to detune the total fluxes threading $O_1P_1$ and $O_2P_2$ being
$(\Phi _0/2)+D$ and $(\Phi _0/2)-D$ respectively, $D>0$. When calculating the
above values of fluxes, the mutual inductance between $O_1$ and $O_2$ is taken
into account but the effects of $J_1$ and $J_2$ are not. Notice that the two
output cells $O_1P_1$ and $O_2P_2$ are always in opposite states, consequently
the back-actions of $O_1$ and $O_2$ to the input cells are canceled out.
Assume that each of the two input ends $J_1$ and $J_2$ contributes the amount
of perturbation flux $+\Delta$ or $-\Delta$ to both the two output heads $O_1$
and $O_2$ according to the input logic value is $0$ or $1$, $\Delta >0$.
Adjust the parameters to satisfy the condition
$$D<2\Delta <\frac{1}{2}\Phi _0 -D$$ 
so that for any logic inputs $I_1$ and $I_2$, the logic value given by
$O_1$ and $O_2$ should be $\overline{I_1\wedge I_2}$ and
$\overline{I_1\vee I_2}$ respectively in order to lower the energy of the
whole system. To see how the thing works, one should bear in mind the fact
that according to the flux threading a superconductive ring due to external
fields is smaller or greater than $\Phi _0/2$, there is supercurrent to lower
the total flux to $0$ or increase it to $\Phi _0$ in order for the ring to be
in the lowest energy state and to satisfy the flux quantization condition at
the same time [4]. Look at the output cell $O_1P_1$, for three inputs
$(I_1,I_2)=(0,0),(0,1),(1,0)$, the total threading flux is always
greater than $\Phi _0/2$ but less than $\Phi _0$, therefore it gives the output
value $1$ that keeps the system in lower energy state. For the only input
$(I_1,I_2)=(1,1)$, the threading flux is less than $\Phi _0/2$ but greater than
$0$, the output should be $0$. Obviously the cell $O_1P_1$ always carries
out the NAND operation. Similarly $O_2P_2$ does NOR. The inverter and the
NAND/NOR gate accomplish the objective logic functions by means of static
quantum logic [2] in the sense that when and only when the whole system is in
the ground state, the states of the output and input rings interpreted
as Boolean variables satisfy the desired logic relations. Using the ``input
symmetrization'' technique [2] one can make all logic gates ``energy
degeneracy conserving'' (EDC) [2]. An ``energy degeneracy lifting'' (EDL)
unit [2] can be implemented by simply applying bias field to detune the
total threading flux off $\Phi _0/2$ thus lead to energy split between the
two wises of supercurrent for the specified cell.

The conclusion is that all the basic elements for the static quantum computer
can be implemented in superconducting circuit. Please notice that the binary
wire approach [1,2,3,7] is not used in this letter because it is
convenient to convey the supercurrent as information carrier through
extended superconducting wires. The technology of
superconductive devices has been well established and widely applied [4,5].
Recently lots of special superconductive devices of scientific interests have
been fabricated to simulate other physical systems or manifest the macroscopic
quantum phenomena, to mention but only a few [6,9,10,11]. Although
implementing a universal static quantum computer may be right within the
reach of today's technology, for the pedagogy purpose at hand it should be
very interesting to construct a simpler, special-purposed machine, such as the
search machine solving the 3SAT problem [1,12]. The proposed 3-clause-evaluator
(3CE) in Ref.[1] uses quantum dots and there are several parameters which
should be carefully tuned. With the superconductive symmetrized-AND (SAND)
gate which is EDC and a DEDLU (decision energy degeneracy lifting unit) [2],
the 3CE for three literals $(l_1,l_2,l_3)$ [1,12] can be easily implemented as
shown in Fig.4.

\begin{center}
{\large FIGURE CAPTIONS}
\end{center}

\noindent
Fig.1\hspace{3 mm}The Ising-spin-ring. The arrows indicate the direction of
supercurrent.

\noindent
Fig.2\hspace{3 mm}A logic inverter with fan-out 2. 

\noindent
Fig.3\hspace{3 mm}The NAND/NOR gate.

\noindent
Fig.4\hspace{3 mm}The 3CE using two SANDs and one DEDLU. 

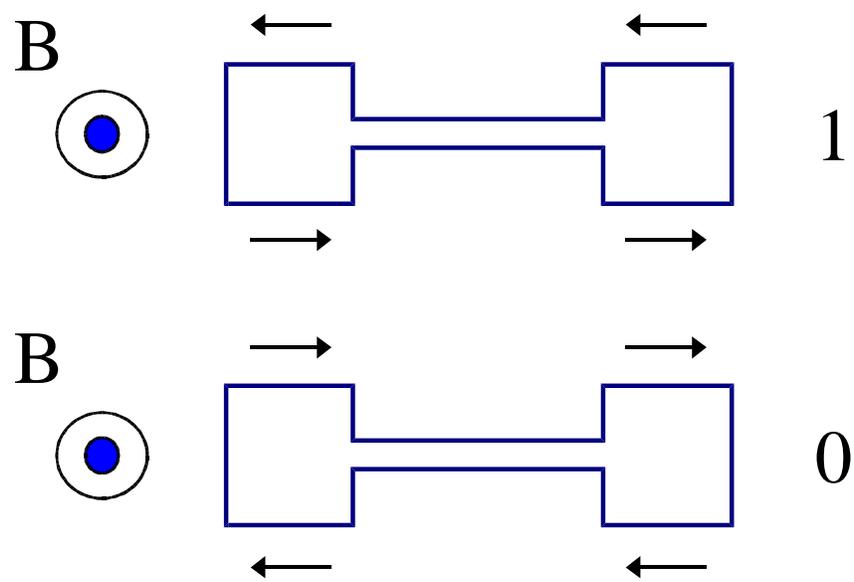
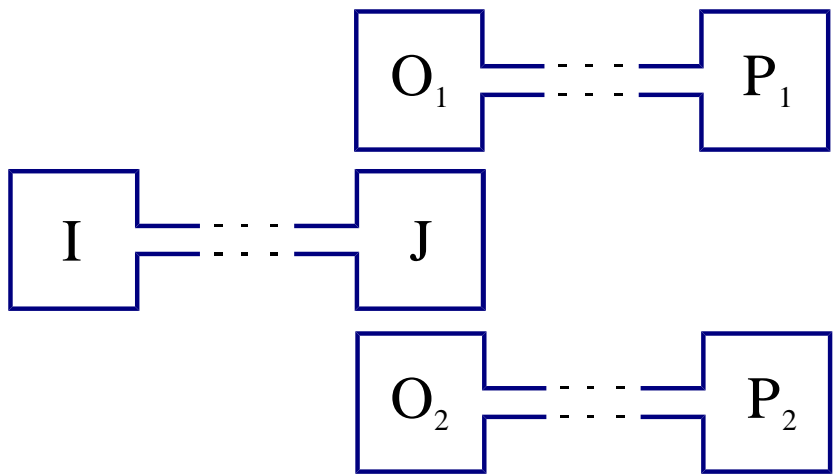
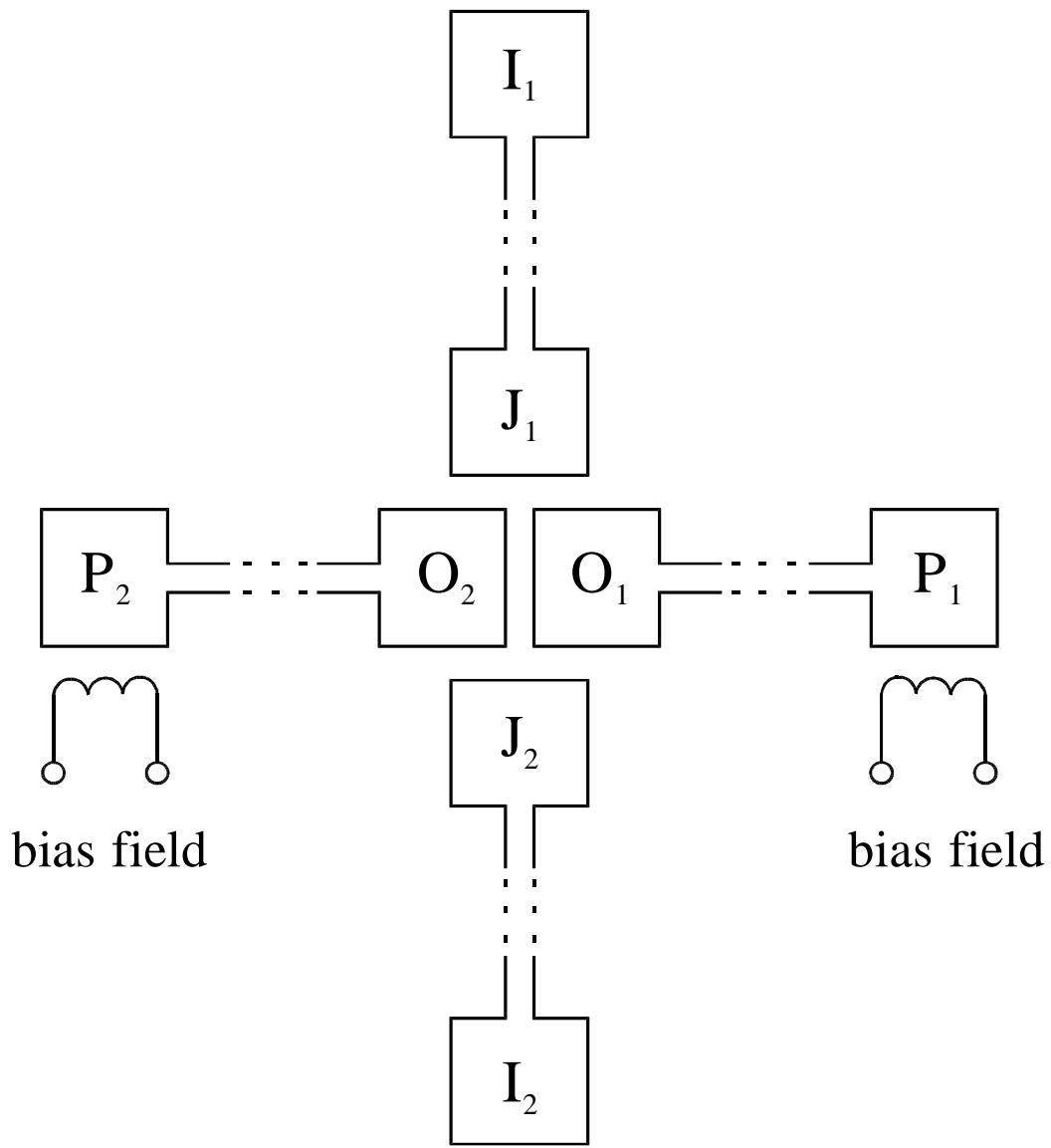
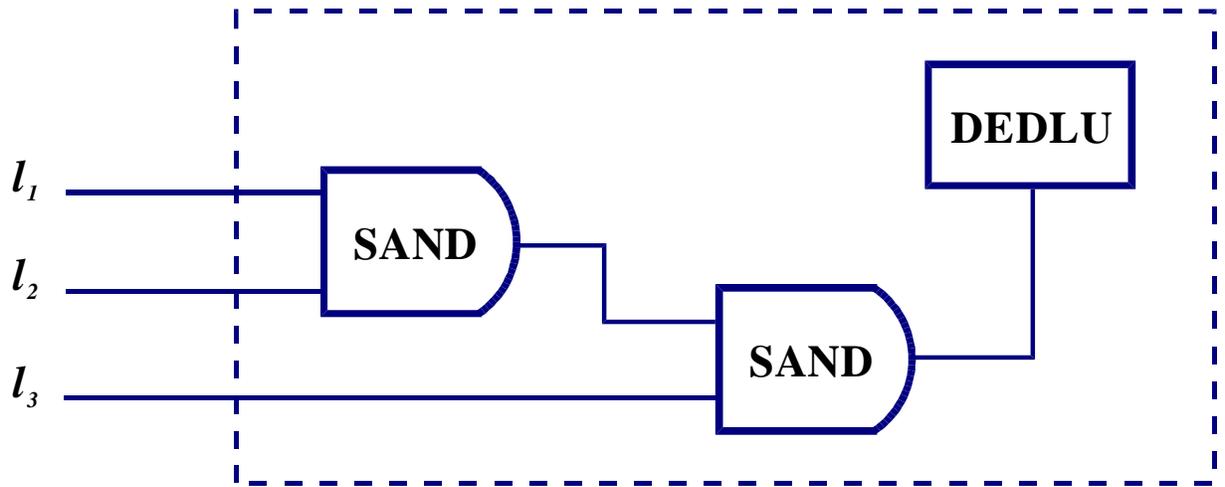
\end{document}